\documentclass[rnote]{aa} 
\usepackage{graphicx}
\usepackage{txfonts}

\def\agt{\mathrel{\raise.3ex\hbox{$>$}\mkern-14mu\lower0.6ex\hbox{$\sim$}}}
\def\alt{\mathrel{\raise.3ex\hbox{$<$}\mkern-14mu\lower0.6ex\hbox{$\sim$}}}

\begin{document}
   \title{Orbital parameters of supergiant fast X-ray transients}

   \subtitle{}

   \author{S. Karino
          \inst{1,2}
          }

   \institute{JAD Program, Universiti Industri Selangor, 
				block 4, Jalan Zirkon A7/A, Seksyen 7, 40000 Shah Alam, Selangor, Malaysia 
         \and
             Shibaura Institute of Technology, Toyosu 3-7-5, Koto, Tokyo 135-8548, Japan \\
              \email{karino@sic.shibaura-it.ac.jp}             
             }

   \date{Received XX YY, 2010; accepted XX YY, ZZZZ}

 
\abstract
{
Super\,giant fast X-ray transient (SFXT) is a new class of the high mass X-ray binary that shows short X-ray flares. The physical mechanism of SFXT short flares is still open for discussion. 
The accretion process of dense clumps in stellar wind onto neutron star (NS) has been proposed as the origin of such short flares.
}
{
In order to examine the applicability of the clumpy wind scenario, we focus on the accretion mode that depends on orbital parameters. Our goal is to impose restrictions on the orbital parameters of SFXT.
}
{
Assuming a simple analytic model of clumpy wind, we investigate the condition where the size of accretion cylinder overcomes the clump size. 
}
{
The allowed parameter region for SFXT is restricted in a relatively narrow window in $P_{\rm{orb}} - e$ diagram. 
}
{
Binary systems with large eccentricities ($e \ga 0.4$) and moderate orbital periods ($P_{\rm{orb}} \sim 10$ d) are prone to show periodic X-ray outbursts which are characteristic for SFXT. 
We confirm that systems with a long orbital period of more than 100 days cannot produce bright X-ray flares in the simple clumpy wind scenario.
}


\maketitle
%

\section{Introduction}

Super\,giant fast X-ray transient (SFXT) is a new class of high mass X-ray binaries (HMXB), found in the monitoring observation of the Galactic plane by INTEGRAL. 
The SFXT is characterized by X-ray outbursts composed of short bright flares, which cease within a few ks. 
It also shows a large dynamic range of X-ray luminosity that reaches 4-5 orders of magnitudes. 
The luminosity of bright X-ray flares exceeds $10^{36} \rm{erg} \, \rm{s}^{-1}$. 
On the other hand, in the quiescent phase, SFXT becomes a dim source whose luminosity is $10^{32} \rm{erg} \, \rm{s}^{-1}$. 
This behavior of SFXT has been reported by Sguera et al. (\cite{S05}), and they are designated as a new 
class of HMXB by Smith et al. (\cite{S06}).
Thereafter, SFXT has attracted the interest of many scientists.

Before the discovery of SFXT, HMXBs were categorized into two classes, the Be-type and the OB-type (Corbet \cite{C84}, \cite{C86}). 
The Be-HMXB is characterized by a Be-type optical counterpart and by a periodic outburst associated with a long orbital period. 
A neutron star (NS) in this system takes a large eccentric orbit around the central Be star. 
The Be star has a dense stellar wind that is confined in its equatorial plane, and at the points where the NS's orbit crosses this equatorial wind, bright X-ray outbursts are observed.
The second classical class of HMXB is a binary system with an OB-type super\,giant (SG) companion. 
Systems categorized in this type are persistent bright X-ray sources, and their orbital periods are much shorter then those of the Be-type HMXBs (typically, a few days). 
Accretion from the SG star to the NS proceeds via isotropic stellar wind or Roche lobe overflow 
in the OB-type systems.
Besides these, the SFXTs show intermediate properties; they show recurrent outbursts and an orbital period between a few days to several 10 days
\footnote{IGRJ11215-5952 has an exceptionally longer orbital period; 
$P_{\rm{orb}} = 165$ days (Romano et al. \cite{R07})
}
 (in't Zand \cite{Z05}; Sguera et al. \cite{S05}). 

The physical mechanism of short X-ray flares of the SFXT is still open for discussion, and several scenarios have been proposed. 
Among them, the most frequently discussed scenario is the so-called clumpy stellar wind model (in't Zand \cite{Z05}; Walter \& Zurita Heras \cite{WZ07}; Leyder et al. \cite{L07}; Rampy et al. \cite{RSN09}). 
In this scenario, an isotropic stellar wind containing a lot of dense clumps is considered. 
These clumps cause short X-ray flares when they accrete onto a NS. 
The clumpy wind accretion scenario can reasonably explain the observed properties of X-ray flares, such as short flare durations, bright X-ray luminosities, large absorptions (in't Zand \cite{Z05}; Walter \& Zurita Heras \cite{WZ07}). 
On the other hand, a few systems can be explained by the different system configuration. 
Sidoli et al. (\cite{S07}) considered a binary configuration that consists of an eccentric NS and a massive star with disk-like anisotropic stellar wind (Sidoli et al. \cite{S07}). 
If the stellar wind is confined to the equatorial plane of the SG, the X-ray outburst arises when the NS passes through the dense stellar wind in the equatorial plane. 
With this scenario, periodic outburst could be naturally explained by associating 
with the system orbital period. 
Another mechanism dominated by the gating effect caused by the NS magnetic and rotational properties has been proposed (Bozzo et al. \cite{B08a}). 
Although this model can explain the large dynamic range of the X-ray luminosity, it is hard to reconstruct a small emission region in quiescent phase. 
Furthermore this model needs an extremely strong magnetic field ($\sim 10^{14}$ G) around the NS, and it is considered to be less pragmatic (Romano et al. \cite{R09c}; Sidoli et al. \cite{S09a}).

In this brief study, we focus on the clumpy wind scenario, and examine how the properties of X-ray outbursts depend on the system's orbital parameters. 
Here, we especially consider the recurrent feature of X-ray outbursts and quiescence that is a typical property of SFXT. 
We investigate the parameter region that allows these recurrent outbursts, assuming a simple clump model. 
By simple calculations, we try to achieve constraints for the system configuration of SFXTs.

\section{Model}

In order to show that the clumpy wind model is the reliable mechanism of the SFXT, we have to check that the considered physical process can explain all observed properties of the SFXT. 
That is, the model should reconstruct the typical orbital period ($\sim 10$ d), the short duration of flare ($\sim 1\,000$ s), the large dynamic range of the X-ray luminosity ($\sim 10\,000$), the dark quiescent phase ($\la 10^{32} \rm{erg} \, \rm{s}^{-1}$ ), the small emission region ($\sim 10^{5}$ cm), the time dependence of the photon index and the absorptions. 
At least for a few systems, the clumpy wind model seems to reconstruct these observed properties very well (for example, IGR J17544-2619; see in't Zand \cite{Z05}; Clark et al. \cite{C09}). 
On the other hand, however, there are serious difficulties 
for few systems (for example, IGR J11215-5952 and IGR J16479-4514)
to explain the observed parameters under the clumpy wind assumption (Romano et al. \cite{R07}; Romano et al. \cite{R09b}; Bozzo et al. \cite{B08b}; Romano et al. \cite{R08}; Jain et al. \cite{J09}). 
Hence, it is considered that the clumpy wind model has a certain limit of application.

The clumpy wind accretion in the SFXT was well considered in Ducci et al. (\cite{D09}). 
In their study, they computed the flare frequency distributions due to clump accretion in broad parameter regions. 
One of the important arguments in their investigations is about the magnitude relation of the clump size ($R_{\rm{acc}}$) and the size of the accretion cylinder (accretion radius; $R_{\rm{acc}}$). 
That is, the accretion mode changes when the indicator of the magnitude relation, $R_{\rm{acc}} / R_{\rm{cl}}$ varies. 
Roughly translated, if the accretion radius exceeds the clump radius, most of the clump mass accretes onto the NS when the clump encounters the accretion cylinder of the NS. 
In this case, a bright X-ray flare occurs because a large amount of potential energy can be liberated. 
On the other hand, if the clump size overwhelms the accretion radius, only a tiny part of the clump mass can be drawn into the accretion cylinder of the NS. 
In this case, bright flares hardly occur even if clumps hit the accretion cylinder because of 
low accretion rates. 
Hence, it is considered that the large accretion radius, which exceeds the clump size, is required as the necessary condition to cause the outburst in accreting systems from clumpy winds.

The existence of X-ray outbursts composed of short flares is one of the most important properties of the SFXT to be explained. 
The condition for outbursts could depend on the magnitude relation, $R_{\rm{acc}}/R_{\rm{cl}}$. 
Here, applying a simple clump model, we examine the ratio $R_{\rm{acc}}/R_{\rm{cl}}$ as the function of orbital parameters. 
If the orbit of NS around SG has a finite eccentricity, this ratio varies with the orbital phase. 
We may consider three situations concerning this magnitude relation: 
(A) if the clump size always exceeds the accretion radius during orbital motion, a bright flare rarely occurs throughout the orbital phase. 
In this case, the system shows no X-ray outburst. 
These systems are dim sources that cannot be detected. 
(B) If the accretion radius always exceeds the clump size throughout the orbital phase, bright flares consistently occur. 
In this case, the system shows persistent outbursts. 
The observed properties of these systems may be rather similar to persistent OB-type HMXBs. 
(C) If the ratio varies and strides across one (namely, the magnitude relation is inverted), a bright flare occurs only in a limited orbital phase. 
In the other phases, the X-ray flare rarely occurs and the system stays in a quiescent phase. 
This transitional behavior between quiescence and outbursts is exactly the feature of the SFXT, hence this inversion of the magnitude relation could be a required condition for an SFXT in the clumpy wind scenario. 
The ratio $R_{\rm{acc}}/R_{\rm{cl}}$ is a function of the orbital phase. 
Then, the orbital parameters of the SFXT should be naturally limited to satisfy the condition (C). 
Hence, we examine the variation of the ratio $R_{\rm{acc}}/R_{\rm{cl}}$ with the orbital parameters, and investigate the condition where $R_{\rm{acc}}/R_{\rm{cl}}$ crosses 1.

Assuming simple formulae for the clumpy wind, we can construct the magnitude relation $R_{\rm{acc}}/R_{\rm{acc}}$ as a function of $r$, the distance from the SG to the NS. 
First of all, the accretion radius is defined as the radius where the kinetic energy of the clump balances the gravitational potential of the NS (in general the kinetic energy is much higher than internal energy in stellar wind; Frank et al. \cite{FKR02}). 
The gravitational potential can be written in the Newtonian framework at the accretion radius
\begin{equation}
\phi = \frac{GM_{\rm{NS}}}{R_{\rm{acc}}} . 
\end{equation} 
As usual, $G$ is the gravitational constant and $M_{\rm{NS}}$ denotes the mass of the NS, respectively.
On the other hand, the kinetic energy of the clump depends on the velocity of the clumpy wind 
($v_{\rm{cl}}$) and the NS orbital velocity ($v_{\rm{NS}}$).
The wind velocity released from the SG stellar surface and accelerated via the line-driven mechanism (Castor et al. \cite{C75}) can be modeled as 
\begin{equation}
v_{\rm{cl}} = v_{\infty} \left( 1 - 0.9983 \frac{R_{\rm{SG}}}{r} \right)^{\beta} , 
\end{equation} 
where $\beta$ is a parameter taking a value between 0.5 and 1.5 (Kudritzki \& Plus \cite{K00}; Walter \& Zurita Heras \cite{WZ07}). 
$R_{\rm{SG}}$ denotes the radius of the SG and $v_{\infty}$ is the terminal velocity of stellar wind, respectively. 
The NS orbital motion can be obtained from the Newtonian equation of motion around a point gravity source (Kepler's laws). 
Under these assumptions, the accretion radius can be obtained as a function of $r$
\begin{equation}
R_{\rm{acc}} = \frac{2 G M_{\rm{NS}}}{v_{\rm{cl}}(r)^2 + v_{\rm{NS}}(r)^2} 
\end{equation} 
(Frank et al. \cite{FKR02}).

On the other hand, the clump size is obtained by the pressure balance between the background stellar wind and the clump itself. 
Romano et al. (\cite{R09a}) and Ducci et al. (\cite{D09}) have used a simple formula to compute the clump size as
\begin{equation}
R_{\rm{cl}} = R_{\rm{cl,0}} \left( \frac{r^2 v_{\rm{cl}}(r) }{R_{\rm{SG}}^{2} v_{\rm{cl,0}} } \right)^{1/3} .
\end{equation}
In this equation, the suffix "0" means the initial values at the SG surface.
We also adopt this simple expression here. 
Then, both the clump size and the accretion radius can be expressed as functions of $r$, the distance from the SG star. 
The position where the clumps encounter the accretion cylinder should be on the orbit of the NS. 
This orbit can be obtained easily via Kepler's laws, if we fix the eccentricity and the orbital period. 
Here, we use normal parameters for the SG and the NS; the mass of the SG is $ M_{\rm{SG}} = 30 M_{\odot} $, the radius of the SG is $ R_{\rm{SG}} = 20 R_{\odot} $ and the mass of the NS is 
$M_{\rm{NS}} = 1.5 M_{\odot} $, respectively. 
As the parameters of the stellar wind, we assume $ R_{\rm{cl,0}} = 2 \times 10^{9}$ cm, $v_{\infty} = 2 \times 10^{8} \rm{cm \, s}^{-1}$ and $\beta = 1.0$.
\footnote{
Here, the used value of $R_{\rm{cl,0}} = 2 \times 10^{9}$ cm is taken from Romano et al. (\cite{R09a}).
The value of $R_{\rm{cl,0}}$ is important, since it directly controls the size relation between accretion radius and clump size. 
However, the value of $R_{\rm{cl,0}}$ is limited as a moderate value.
In Ducci et al.~(\cite{D09}), the maximum and minimum values of the clump size are discussed.
The clump should be small enough to be denser than the background wind, while the minimum size of the clump is limited by the optical thickness condition.
The possible size of $R_{\rm{cl,0}}$ depends on the clump mass, $M_{\rm{cl}}$, and if we assume $M_{\rm{cl}} \sim 10^{20}$ g, the initial clump radius is $2 \times 10^9 {\rm cm} \la R_{\rm{cl,0}} \la 2 \times 10^{10} {\rm cm}$ (see Fig.~3 in Ducci et al.~\cite{D09}).

If we select a low (high) value of $R_{\rm{cl,0}}$, flares easily (hardly) occur and the SFXT region in Fig.~1 (see Section 3) becomes larger (smaller).
However, since the allowed range of $R_{\rm{cl,0}}$ is not very broad especially for large $M_{\rm{cl}}$, the change of the SFXT region is moderate. 
}

\section{Results}

Applying these formulae of the clump size and the accretion radius, we can compute the ratio $R_{\rm{acc}} / R_{\rm{cl}} $ for various values of the orbital period $P_{\rm{orb}}$ and the orbital eccentricity $e$. 
For each set of $(P_{\rm{orb}}, e)$, we compute the variation of the ratio $ R_{\rm{acc}} / R_{\rm{cl}} $ through the orbital motion. 
As a result, the systems can be divided into three categories by the variation modes of this ratio: case (A) is that the ratio is always lower than 1; case (B) is that the ratio always exceeds 1; and case (C) is that the ratio varies and is equal to 1 at certain orbital passages. 
As discussed above, the first and second cases correspond to quiescent and persistent sources. 
Only the systems categorized in the third case can reconstruct the observed behavior of the SFXT. 
This result is summarized in Fig.~1.

In Fig.~1 the left-side region indicated by small crosses ($+$) denotes the prohibited region for binaries, because the NS orbit enters the interior of the SG. 
In this region, the NS may suffer from strong dragging force via interaction with the SG matter, and cannot stay in the stable orbit. 
Perhaps the orbits of these systems shrink quickly and evolve toward coalescence (Rosswog \& Br\"{u}ggen \cite{RB07}).

The right-lower broad region filled with black dots ($\cdot$) is a parameter region where the clump size exceeds the accretion radius throughout the orbital motion (case A). 
As discussed above, in this situation, only a part of clump mass can accrete onto the NS, even if the clump encounters the NS. 
Consequently, the system in this region cannot produce bright flares. 
These systems become dim X-ray sources, which emit via background wind accretion.

The left-lower shaded part in the figure indicates the region where the accretion radius exceeds the clump size throughout the orbital motions (case B). 
In this case, collisions of clumps and the accretion cylinder consistently produce bright flares. 
At the same time, since the NS orbit is near to the SG surface, the number density of clumps is very high in these systems. 
It means that the frequency of clump collisions onto the accretion cylinder is also high. 
Consequently, systems in this region always show outbursts and are observed as persistent X-ray sources. 
This situation may correspond to the classical HMXB with an OB companion.

The remaining central-upper region labeled gSFXTh corresponds to case (C). 
Systems in this region show the inversion of the magnitude relation between clump size and accretion radius during orbital motion. 
Only systems in this region have both quiescence and outburst through the orbital motion of the NS. 
In these systems, the NS does not show bright X-ray flares around the apastron passage (quiescent phase). 
On the other hand, near the periastron, the accretion radius overwhelms the clump size and it produces bright flares (outburst phase). 
As a crude summary, according to Fig.~1, only systems with (i) an intermediate orbital period of $\sim 10-30$ d, and/or (ii) large eccentricity could reconstruct the burst properties of the SFXT.
Because the SFXT window that we derived above is somewhat narrow in the parameter space, it makes a strong restriction to the orbital parameters of the SFXT systems.

Next, we compare actually observed SFXT systems with our result. 
At present, $\sim$ 15 candidate systems of SFXT have been found (Bodaghee et al. \cite{B07}; Walter \& Zurita Heras \cite{WZ07}). 
Among them, we list the systems whose orbital parameters have been well studied in Table~1. 
Additionally, we plot the orbital parameters of these SFXTs by filled circles with labels;
(a)IGR J16479-4515, (b)IGR J17544-2619, (c)IGR J18483-0311, (d)SAX J1818.6-1703, (e)IGR J16465-4507, and (f)XTE J1739-302.
For these systems, the eccentricities still have large uncertainties.
Hence we show the error bars for systems (c) and (d) (the sizes of error bars are taken from Rahoui \& Chaty \cite{RC08} and Zurita Heras \& Chaty \cite{ZC09}, respectively).
For other systems, only lower limits (for systems (a) and (b); see Romano et al. \cite{R08} and Clark et al. \cite{C09}) or upper limits (for systems (e) and (f); see Clark et al. \cite{C10} and Drave et al. \cite{D10}) are known.
Because the orbital periods of these systems are extensively studied, the error bars in the horizontal direction are small.
The orbital period of (g)IGR J11215-5952 is 165 [d] and this value is situated far outside of the boundary box of this figure.
Unfortunately, the eccentricities of IGR J08408-4503 and IGR J11215-5952 are still unknown.

For systems (b), (c), (e) and (f), the observed orbital parameters seem to be almost consistent with our theoretical model, although (b)IGRJ17544-2619 is just located on the border of our SFXT window. 
Additionally, it is said that also IGRJ08408-4503 ($P_{\rm{orb}} = 35$ d, $e$ is unknown) has large eccentricities (Leyder et al. \cite{L07}; Romano et al. \cite{R09a}; Bird et al. \cite{B09}; Sidoli et al. \cite{S09b}). 
Although precise measurements have not been reported, this system may enter the SFXT window, or locate around the boundary. 
From Figure 1 it seems that the eccentricity of this system could be quite large ($e$ should be larger than 0.5). 
The observational confirmation of the eccentricity is required.


On the other hand, (d)SAXJ1818.6-1703 seems to stray outside the SFXT window. 
However, Zurita Heras \& Chaty (\cite{ZC09}) obtained its eccentricity as $\sim 0.4$ by simple assumptions of the system configuration. 
In their study, they assumed that X-ray outburst occurs only when the NS enters in the domain of $2 R_{\rm{SG}} < r < 3 R_{\rm{SG}}$. 
If we consider the magnitude relation $ R_{\rm{acc}} / R_{\rm{cl}}$, a lager eccentricity may be required to cause outburst. 
%
Note that our assumption that $R_{\rm{acc}} / R_{\rm{cl}} = 1$ decides the SFXT boundary in $P_{\rm{orb}} - e$ plane is not strict. 
The boundary could have a certain level of ambiguity. 
Even if $R_{\rm{acc}} < R_{\rm{cl}}$, a head-on collision between the clump and the NS accretion cylinder could produce a high accretion rate and consequently bright flares.
Considering the conditions above, we may regard the location of SAXJ1818.6-1703
as a marginal system between an SFXT and a quiet source in Fig.~1.

\bigskip

Here, we have made the quite simple assumption that only the $R_{\rm{acc}} / R_{\rm{cl}}$ ratio decides the outburst conditions of X-ray binaries. 
However, our result reconstructs roughly the thresholds of the SFXT, although there are a few exceptional systems, which we discuss below. 
In the estimates of the SFXT parameters from limited observations, a guideline like this will be beneficial.


\begin{figure}
\centering
\includegraphics[angle=0,width=8.5cm]{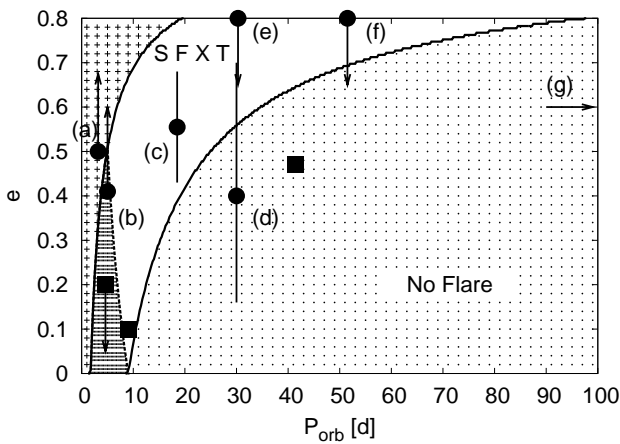}
\caption{
SFXT window in $P_{\rm{orb}} - e$ plane. 
The region filled with small $+$ denotes the prohibited region where binaries cannot stay in stable orbits.
Systems in the right-lower region filled with black dots produce no X-ray flares, since $R_{\rm{acc}} < R_{\rm{cl}}$ through the orbital motion (case A). 
The left-lower shaded part indicates the region where $R_{\rm{acc}} > R_{\rm{cl}}$ through the orbital motion (case B). 
Systems in this region will show persistent outburst. 
The central-upper region denotes SFXT window characterized by quiescence and outburst (case C). 
Filled circles show the position of the observed SFXTs; (a)IGR J16479-4515, (b)IGR J17544-2619, (c)IGR J18483-0311, (d)SAX J1818.6-1703, (e)IGR J16465-4507, (f)XTE J1739-302.
The orbital period of (g)IGR J11215-5952 is 165 [d] and this object is situated far outside of this figure.
The filled square indicates observed classical HMXBs; 
SAX J1802.7-201 ($P_{\rm{orb}}=4.6$[d], $e < 0.2$),
Vela-X1 ($P_{\rm{orb}}=9$[d], $e = 0.1$) and GX301-2 ($P_{\rm{orb}}=41.5$[d], $e = 0.47$).
}
\label{fig1}
\end{figure}

\begin{table*}
\caption{List of the SFXT systems whose orbital parameters have been studied. 
Some classical HMXBs are also shown for comparison.}             
\label{table:1}      
\centering          
\begin{tabular}{c c c c c c}      
\hline\hline       
name & label in Fig.1 & $P_{\rm{orb}}$[d] & eccentricity & comment & reference$^{*}$ \\
\hline
IGR J16479-4514 & (a) & 3.2 & $> 0.5$ & eclipse? & (1) (2) (3) (4) \\
IGR J17544-2619 & (b) & 4.9 & $> 0.4$ & eclipse? & (5) (6) (7) \\
IGR J18483-0311 & (c) & 18.5 & $0.43 - 0.68$ & eclipse? & (8) (9)  \\
SAX J1818.6-1703 & (d) & 30 & $\sim 0.4$ & & (10) (11) (12)\\
IGR J16465-4507 & (e) & 30.3 & $< 0.8$ & sgHMXB? & (13) (14) \\
IGR J08408-4503 & & 35 & & & (15) (16) \\
XTEJ1739-302 	& (f) & 51.5 & $< 0.8$ & & (17) (18) (19) \\
				& & (12.9?) & & & \\
IGR J11215-5952 & (g) & 165 & & Disc-like wind accretion? & (20) (21) \\
\hline
SAX J1802.7-201 & & 4.6 & $< 0.2$ & classical HMXB & (22) (23) \\
Vela X-1 & & 9 & 0.1 & classical HMXB & (24) (25) \\
GX301-2 & & 41.5 & 0.47& classical HMXB & (26) (27) \\
\hline                  
\end{tabular}

\begin{list}{}{}
\item[$*$] 
(1)Bozzo et al \cite{B08b}; (2)Romano et al. \cite{R08}; (3)Sguera et al. \cite{S08}; 
(4)Jain et al. \cite{J09}; 
(5)in't Zand \cite{Z05}; (6)Clark et al. \cite{C09}; (7)Rampy et al. \cite{RSN09}; 
(8)Rahoui \& Chaty \cite{RC08}; (9)Romano et al. \cite{R10}; 
(10)Bird et al. \cite{B09}; (11)Sidoli et al. \cite{S09b}; (12)Zurita Heras \& Chaty \cite{ZC09}; 
(13) Clark et al. \cite{C10}; (14) La Parola et al. \cite{L10}
(15)Leyder et al. \cite{L07}; (16)Romano et al. \cite{R09a}; 
(17)Romano et al. \cite{R09c}; (18)Sidoli et al. \cite{S09a}; (19)Drave et al. \cite{D10}; 
(20)Romano et al. \cite{R07}; (21)Romano et al. \cite{R09b}; 
(22)Augello et al. \cite{A03}; (23)Hill et al. \cite{H05};
(24)White et al. \cite{W83}; (25)Nagase \cite{N89}; 
(26)Watson et al. \cite{W82}; (27)Koh et al. \cite{K97}

\end{list}

\end{table*}

\section{Discussions}

In the list of observed SFXTs (Table 1), IGRJ16479-4514 has an extremely short orbital period, $P_{\rm{orb}} = 3.2$ d (Jain et al. \cite{J09}). 
Although it is suggested that this system has a large eccentricity, such a system with short orbital period likely becomes a persistent source or contact binary, according to our result (see Fig.~1). 
It is also suggested that the dark quiescent phase cannot be explained in the wind accretion model for such a small $P_{\rm{orb}}$ system (Romano et al. \cite{R09c}). 
Hence, it is difficult to understand the observed X-ray properties under the clumpy wind scenario, and it may need other models; e.g. a disk-like wind model (Romano et al. \cite{R09c}; Sidoli et al. \cite{S09a}). 
Some authors have suggested that the eclipse by the SG star causes the X-ray dark phase in this system (Bozzo et al. \cite{B08b}). 
Since this system shows marginal properties between the classical OB-type HMXB and an SFXT, it could be a prototype system that bridges new and classical understandings of HMXBs (Sguera et al. \cite{S08}; Jain et al. \cite{J09}). 
Further observations and discussions on this source are strongly required.

Meanwhile in IGRJ11215-5952, the observed orbital period ( $P_{\rm{orb}} = 165$ d ) is much longer than in other systems. 
From our result based on the magnitude relation $R_{\rm{acc}}/R_{\rm{cl}}$ in the clumpy wind scenario, the eccentricity of such long $P_{\rm{orb}}$ SFXTs should be almost 1. 
Because this large eccentricity is not realistic, it may be natural to consider other mechanisms for this system. 
To explain its precise periodicity and X-ray outburst properties, a disk-like wind mechanism has been examined for this system (Sidoli et al. \cite{S07}; Romano et al. \cite{R09b}). 
Fortunately, the spin period of the NS has been measured in this system (Romano et al. \cite{R07}). 
The set of parameters, ($P_{\rm{orb}} = 165 {\rm {d}}, P_{\rm{spin}} = 187 {\rm{s}}$) in this system locates exactly on the Be-HMXB band in Corbet diagram (Corbet \cite{C84}, \cite{C86}).
Since the outburst mechanism proposed by Sidoli et al. (\cite{S07}) is almost the same as classical Be-HMXBs, we can speculate that this system could be a special type of a Be-type HMXB. 
That is, perhaps we observe this system from a face-on angle, and the extinction by a disk-like wind that is confined in the equatorial plane cannot be observed.

In Fig.~1 some selected classical HMXBs are also shown with a filled square
(see also Table~1).
These systems are SAX J1802.7-201 ($P_{\rm{orb}}=4.6$[d], $e < 0.2$),
Vela-X1 ($P_{\rm{orb}}=9$[d], $e = 0.1$) and GX301-2 ($P_{\rm{orb}}=41.5$[d], $e = 0.47$), respectively.

SAX J1802.7-201 is a classical persistent HMXB, which contains an OB giant with  
$R \sim 20 R_{\odot}$ and $M \sim 25 M_{\odot}$ (Hill et al. \cite{H05}).
Hence, its stellar parameters agree well with our model assumptions.
Actually, this system is situated exactly in the persistent HMXB region in Fig.~1.

Among three samples of HMXB, the well known object Vela X-1 is somewhat problematic.
Its orbital parameters are within the SFXT region (Fig.~1), although Vela X-1 has been known as a persistent bright X-ray source.
Since this system is located near by the boundary of the SFTX,  
Vela X-1 may be considered as the intermediate object between an SFXT and a classical HMXB.
Recently, short X-ray flares from Vela X-1 have been detected (Kreykenbohm et al. \cite{K08}).
These flares seem similar to that of SFXTs, however, Vela X-1 critically differs from SFXTs by the lack of off-states.
The flare behavior of this object may be understood as follows;
in Vela X-1 as well, bright and short bursts could be made by clump accretions.
However, even without clump accretion, back\,ground wind could maintain a bright, persistent X-ray luminosity, because Vela X-1 has a short orbital distance.
These distinctions of binary classes and the studies of intermediate objects are the hot issues in this field (Sidoli et al. \cite{S10}).

The classical HMXB GX301-2 has a long orbital period ($P_{\rm{orb}}=41.5$[d]) and a large eccentricity ($e = 0.47$) (Watson et al. \cite{W82}).
This system, shown by a filled square in the gNo Flareh region, is another peculiar object.
Although it cannot show X-ray outbursts under our clumpy wind scenario, it is observed as a bright persistent X-ray source.
Additionally, the accretion mode of this system is also peculiar.
In this system, the wind accretion proceeds besides a streaming accretion (Leahy \& Kostka \cite{L08}).
Curiously, moreover, recent observation detected an unusual X-ray flare from GX301-2 during its apastron phase (Finger et al. \cite{F10}).
Some of these strange observational properties may be explained by the hugeness of the giant companion.
The SG star in this system is quite large ($M \sim 50 M_{\odot}$ and $R \sim 60 R_{\odot}$), so our assumption used in Fig.~1 cannot be directly applied (Koh et al. \cite{K97}; Kaper et al. \cite{K06}).

Additionally we have to note that Fig.~1 is constructed under the assumption of the clumpy wind accretion scenario.
This scenario seems somehow reasonable for the SFXTs, however, it is not trivial that the clumpy wind accretion is the general manner in a HMXB.
If any accretion modes other than the clumpy wind proceed, the position of these HMXB systems may not be restricted by our model in Fig.1 (see GX301-2).
In other words, if a persistent source (classical HMXB) is located in the left-lower region in Fig.~1, clumpy wind accretion may be the dominant mechanism in this system the (for instance, SAXJ1802).

\bigskip 

In this study, we discussed the possibility that bright flares occur based only on the magnitude relation between clump size and accretion radius. 
In a practical sense, the occurrence frequency of flares depends on other conditions as well. 
Especially the frequency of actual encounters of clumps with the accretion cylinder is importantD
The encounter frequency of clumps with the NS is also related to the magnitude relation of  $R_{\rm{acc}}$ and $R_{\rm{cl}}$. 
A detailed analysis of the occurrence rate of flares has been given in Ducci et al. (\cite{D09}). 
Concerning the orbital parameters, they show that 
(1) when the orbital period becomes longer, the number of bright flare decreases; 
(2) when the eccentricity becomes larger, the number of bright flare increases (see Fig. 9 in Ducci et al. 2009). 
Our result shown in Fig.~1 basically reconstructs their results.

In Fig. 2 we show the accretion radius and clump size as functions of orbital phase.
We focus on the systems with $P_{\rm{orb}} = 15$ d because these systems easily become SFXT
\footnote{
Indeed, systems likely behave as SFXT when $P_{\rm{orb}} = 10$ d in Fig~1. 
However, by checking the case of $P_{\rm{orb}} = 15$ d, we can confirm that actually the accretion radius is too small when the eccentricity is low ($e=0.1$).
}.
In this figure, the solid curve denotes the normalized solid angle of the accretion cylinder observed from the center of the SG ($R_{\rm{acc}}^{2} / r^{2}$), while the dashed line shows that of the clump ($R_{\rm{cl}}^{2} / r^{2}$). 
The former corresponds to the size of the "target" for clumps (in this analogy, the clump size corresponds to the size of the "bullet"). 
Clearly, a bullet easily hits the target when the size of the target is large. 
In the figure, four solid curves indicate different eccentricities from 0.1 to 0.7, respectively. 
The results show that the accretion radius drastically varies with the orbital phase. 
It takes a huge solid angle near the periastron (phase 0.5), while it shrinks rapidly approaching apastron (phase 0 and 1). 
Its dynamic range achieves more than 3 orders of magnitude in the most eccentric system. 
On the other hand, the clump size varies only mildly. 
This suggests that the variation of the encounter rate between clumps and the accretion cylinder mainly depends on the variation of the target size.

When the NS reaches periastron (phase 0.5), the solid angle of the accretion cylinder takes maximally $10^{-3}$ of the celestial sphere (its apparent diameter becomes $\sim$ 5 degree if we observe from the SG). 
Assuming that clumps are released isotropic manner, a large fraction of clumps can hit the target because the solid angle of the target is large enough. 
As we discussed, in this situation with large $R_{\rm{acc}}$, the accretion of the clump produces bright flares because entire mass of the clump can accrete onto the NS. 
Hence, we can conclude that around periastron in eccentric systems, X-ray outbursts composed with many flares are likely to be observed.
On the other hand, when the NS is far from the SG, the solid angle of the accretion cylinder is quite small. 
Because it is only a tiny target for the clumps to hit, the encounter of the clump with the accretion cylinder is a rare event around apastron. 
Additionally, even if a clump hits the accretion cylinder, a bright flare cannot occur because only a small fraction of the clump mass can accrete onto the NS when 
$R_{\rm{acc}} \ll R_{\rm{cl}}$. 
Consequently, it is clear that no X-ray flare occurs around orbital apastron and the system should be quiescent in eccentric systems.

\begin{figure}
\centering
\includegraphics[angle=0,width=8.5cm]{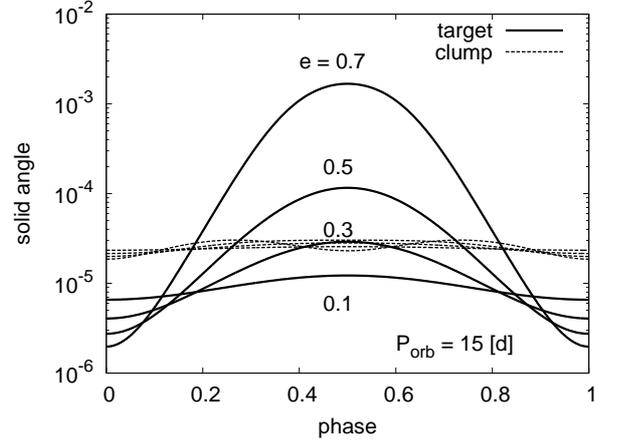}
\caption{
The solid curves denote the accretion radii
for different eccentricities ($e=0.1, 0.3, 0.5, 0.7$).
Dashed curves show the variations of the clump sizes in the same systems.
Phase 0 corresponds to apastron passages.
The orbital period is fixed as $P_{\rm{orb}} = 15$ d.
With this $P_{\rm{orb}}$, the clump size is larger than that of the accretion radius when the eccentricity is low.
Hence, flares rarely occur for these low $e$ cases.
On the other hand, when $e$ is higher, the size of accretion radius becomes larger than clump size and the outburst may occur.
This size-relation is reflected in Fig.~1.
}

\label{fig2}
\end{figure}

\bigskip

At present, studies of clumpy wind accretion scenario have been based on very simple clump models (Walter \& Zurita Heras \cite{WZ07}; Ducci et al. \cite{D09}). 
Of course, however, the actual clumps in the stellar wind may show complex behaviors. 
For instance, the clump size depends on the instability of the SG atmosphere, and is not uniform (Runacres \& Owocki \cite{RO02}). 
$R_{\rm{cl}}$ may obey fairly complex distributions depending on the stellar conditions. 
Additionally, in contrast to the background thin wind whose internal energy is negligible, clumps have a high density and their fluid properties cannot be neglected. 
Neglecting fluid dynamic effects, instabilities, collisions between clumps, and effects of line absorptions, we may overlook some important phenomena and miss essential issues (Howk et al. \cite{H00}; Runacres \& Owocki \cite{RO02}; Oskinova et al. \cite{O07}).

Although it is quite difficult to solve all those problems, a complete understanding of the SFXT may yield even more astronomical benefits. 
Super\,giant stars in the SFXT are very young ($T < 10^{6}$ yr).
If we can obtain information about evolutional passes of massive stellar systems via SFXT observations, these could be great hints about the mysterious processes of massive star-formation. 
Additionally, binaries containing an NS and a massive star, including SFXTs could be progenitors of NS-NS or NS-BH binaries. 
The NS-NS binaries are likely candidates for short gamma-ray bursts, hence now they are studied actively in both theoretical and observational methods. 
In order to obtain the initial condition of these NS-NS binaries, it may be useful to investigate the evolution of SFXTs and classical HMXBs (Rosswog  \& Br\"{u}ggen \cite{RB07}).

\bigskip

In this study, we investigated the magnitude relation between clump size and accretion radius throughout the orbital motion in an SFXT system, applying the simple clumpy stellar wind model. 
Under the simple assumption that the boundary of the SFXT window is indicated by the curve $R_{\rm{acc}} / R_{\rm{cl}} = 1$, we divided the parameter space into three subgroups; persistent source, quiescence, and SFXT. 
In our result, the allowed parameter region becomes a relatively narrow window in a $P_{\rm{orb}} - e$ diagram, which means that our result gives strong restrictions on the orbital parameters of SFXT systems. 
Actually, at least for several SFXT systems whose orbital parameters have been known, their location in Fig.~1 is consistent with the obtained SFXT window. 
On the other hand, we confirm that the systems with a long orbital period of more than 100 days cannot produce bright X-ray flares due to clumpy wind accretion. 
Hence, for SFXT systems with long $P_{\rm{orb}}$, another mechanism, such as the disk-like wind model, should be considered. 
The criterion shown in Fig.~1 is useful to find this peculiarity of X-ray binary systems.
These peculiarities also suggest that there may be greater diversities we have not yet understood in HMXB systems.
Further observations and statistical discussions are required to grasp the true nature of SFXT systems.

\end{document}